\documentclass[structabstract]{aa}  
\usepackage{amssymb}
\usepackage{amsmath}
\usepackage{txfonts}
\usepackage{subfig}
\usepackage{graphicx}
\usepackage{rotating}
\usepackage{url}
\usepackage{epsfig}
\usepackage{color}
\usepackage{supertabular}
\usepackage{array}

\def\gr{$\gamma$-ray}

\usepackage{natbib}
\bibpunct{(}{)}{;}{a}{}{,}

\begin{document}

\date{\empty}

\title{Timing properties of gamma-ray bursts detected by SPI-ACS detector on  board of INTEGRAL}

\institute{     ISDC Data Centre for Astrophysics, Ch. d'Ecogia 16, 1290 Versoix, Switzerland,\\
  Observatory of Geneve, University of Geneva, Ch. des Maillettes 51, 1290 Sauverny, Switzerland       }
   \date{Received ; accepted }

\author{V.Savchenko, A.Neronov. T.J.-L.Courvoisier}
\titlerunning{Timing properties of gamma-ray bursts detected by SPI-ACS detector}

\date{Received $<$date$>$  ; in original form  $<$date$>$ }


\label{firstpage}

\abstract {} { We study timing properties of a large sample of \gr\ bursts (GRB)
  detected by the Anti-Coincidence Shield (ACS) of the SPI spectrometer of
  INTEGRAL telescope.  } { We identify GRB-like events in the SPI-ACS data. The
  data set under investigation is the history of count rate of the SPI-ACS
  detector recorded with a binning of 50~ms over the time span of $\sim 10$~yr. In spite of the fact that SPI-ACS does not have imaging capability, it provides high statistics signal for each GRB event, because of its large effective
 area.  } { We classify all isolated excesses in the SPI-ACS count
  rate into three types: short spikes produced by cosmic rays, GRBs and Solar flare induced events. We find some $\sim 1500$ GRB-like events in the 10 yr
  exposure. A significant fraction of the GRB-like events identified in SPI-ACS
  occur in coincidence with triggers of other gamma-ray telescopes and could be
  considered as confirmed GRBs. We study the distribution of durations of the
  GRBs detected by SPI-ACS and find that the peak of the distribution of long
  GRBs is at $\simeq 20$~s, i.e. somewhat shorter than for the long GRBs
  detected by BATSE. Contrary to the BATSE observation, the population of short
  GRBs does not have any characteristic time scale. Instead, the distribution of
  durations extends as a powerlaw to the shortest time scale accessible for
  SPI-ACS, $\le 50$~ms. We also find that a large fraction of long GRBs has a
  characteristic variability time scale of the order of $1$~s. We discuss
  the possible origin of this time scale.  } {}


\maketitle

\section{ Introduction }\label{intro}

Gamma-ray bursts (GRBs) are the most violent explosions in the
 universe. They are among the first cosmic \gr\ sources discovered more than $40$
 years ago \citep{thegrb}. In spite of the long observation history, the nature of the GRB
phenomenon remains obscure. A number of missions were dedicated to the
study of the physical properties of GRBs. The isotropy of GRB distribution, indicating
their extragalactic origin, was established by CGRO/BATSE observations
\citep{batse4b}. Long-lasting afterglows discovered by Beppo/SAX \citep{agfirst}
confirmed the cosmological nature of GRB phenomenon and provided an insight in
the energy budget of GRBs comparable to the energy release in the core collapse
of massive stars. Complex time structure of GRB lightcurves, with variability
down to millisecond time scale was revealed by CGRO/BATSE
\citep{grbms}. Existence of an additional delayed emission component in the GeV
energy band was discovered by CGRO/EGRET \citep{egretgrb} and was recently
confirmed by Fermi/LAT \citep{grb090902b}.

The transient nature of GRBs makes the information available about each
individual event usually very fragmentary: a complete picture of the GRB
phenomenon could not be established based on the observations of individual
events. Instead, the complete picture could be obtained via population study in
which different events appear as manifestations of a single GRB phenomenon.  The
population study should in the first place clarify if all the observed GRBs are
produced by one and the same physical phenomenon or if the set of GRBs consists
of several sub-populations which might possibly have different physical origin.
This problem is not simple: the quality of the data on more numerous weak bursts
is not sufficient for quantitative analysis, while the bright bursts are not
frequent enough to form a good statistical sample.

Up to now, the largest sample of GRBs available for a statistical study has been
collected by the CGRO/BATSE \citep{batse4b} and Konus-Wind
\citep{konuswind}. One of the main findings of BATSE and Konus-Wind was the
discovery of a dichotomy in the distribution of GRB durations
\citep{shortandlong,grbtwoclasses}. This distribution reveals two broad
bumps. The first bump is at the durations $T_{90}< 2$~s (where $T_{90}$ is the
time over which 90\% of the GRB fluence is registered), corresponding to the
"short" GRBs. The second bump is at $T_{90}>2$~s, corresponding to the "long"
GRBs. In the BATSE GRB sample, the population of long GRBs peaked at
$T_{90}\simeq 50$~s, while the typical time scale of the short GRBs was
$T_{90}\simeq 0.3$~s \citep{grbtwoclasses}. Further analysis of a
  larger sample of BATSE bursts using improved duration estimates found that
  the durations of the long bursts seem to peak in $30-40$~seconds range rather
  then at $50~$seconds \citep{batse4b}

The division of GRB in two sub-populations is supported by observation of
differences in spectral properties of the two sub-classes: the short GRBs have
harder spectra \citep{grbtwoclasses}. Long GRBs are characterized by larger
spectral lags \citep{sgrblags}.  The spatial locations of the bursts from the
two sub-classes in their host galaxies are also different: long GRBs are tracing
the star formation while the short GRBs are not correlated with the
star formation regions \citep{shortgrborigin}.

The short versus long dichotomy is commonly believed to be related to the
difference in burst progenitors. The long GRBs seem to be associated with the
final stage of evolution of massive stars: gravitational collapse of the stellar
core and/or supernova explosion \citep{grbcollaps}.  The short GRBs are possibly
related to coalescence of compact objects - neutron stars or black holes
\citep{shortgrborigin}. In a number of cases supernovae associated with long GRBs
\citep{grbsnrev} were directly observed. Direct confirmation of the hypothesis of
association of the short GRBs to the merger events is still missing.

Some problems of the division of the entire GRB populaiton in two classes remain
\citep{Zhang11_grbstatus}. In particular, tight upper limits on the optical flux
of several long GRBs exclude supernova as the GRB progenitor
\citep{grbnosn,grbnosn2,grbnosn3}.  A peculiar long GRB with firmly excluded
supernova progenitor, GRB 060614A \citep{grb060614} has also negligible spectral
lag typical for the short bursts.  It is also possible that a separate population of
softer spectra GRB-like events exists, the X-ray flashes (XRF). It is, however,
not clear if the XRFs are just a softer tail of the GRB distribution \citep{xrf}
or whether they are a different phenomenon \citep{thexrf}. Large fraction of
short GRBs may contain softer extended emission component, making their total
duration as large as tens of seconds \citep{shortextbatse}. In such situation
the possibility of division of the GRB population in "short" and "long"
sub-classes becomes problematic. As claimed by \cite{vsb}, the shortest bursts,
with durations of less than $100$~ms, may also posses remarkably distinct
properties: unusual spectra and non-isotropic angular distribution
(other studies of anisotropy of GRB subpopulations did not, however,
  come to the same conclusion \citep{grbisotrop}), a luminosity distribution
suggesting Galactic origin.  A population of Galactic sources, soft
gamma-repeaters (SGR), is known to produce short GRB-like events
\citep{sgr1820natur,Hurley09,savchaxp}. The brightest of these events could be
detected by existing instruments even if they occur outside the Milky Way, in
the nearby galaxies. This type of events could constitute a non-negligible
fraction of the observed short GRB sample \citep{sgr1820natur,Hurley09}.

The focus of the present study is on the timing properties of GRBs detected by
the Anti-Coincidence Shield (ACS) of spectrometer SPI on board of the INTEGRAL
sattellite \citep{acs}. Although SPI-ACS does not have an imaging capability, it
has a large effective area (up to about $1$~m$^2$). SPI-ACS readout does not
rely on any on-board trigger criteria and instead continuously measures the
event rate all over the INTEGRAL mission lifetime with a time binning of $50$~ms.

The first attempt to systematically study GRBs with SPI-ACS was done by
\citet{rau05} using about 800 days of data. This study considered 179 GRBs
also detected by other satellites, but visible in the SPI-ACS. Unlike
\citet{rau05}, we primarly rely on identification of the GRBs based solely on
the ACS data, so that our GRB sample is not dependent on the data of other
satellites. This allows us to perform a study of the timing properties of GRBs
which is not affected by the selection biases (sensitivity, trigger
configurations) of other telescopes.  We separately discuss a sub-sample of GRBs
confirmed by the other satellites (Konus/Wind, Fermi/GBM, RHESSI, Suzaku/WAM,
Swift/BAT, INTEGRAL/ISGRI, AGILE, IPN network).

In the following sections we describe in detail the algorithm of detection of
bursts in the SPI-ACS data and the classification of events into three classes:
GRBs, instrumental short spikes and Solar flares. For each GRB we measure its
duration and investigate if the GRB lightcurve possesses a characteristic time
scale other than the overall duration. Based on this study we build the
distribution of GRB durations and characteristic time scales.

The main obstacle for the study of GRBs with SPI-ACS is the presence of the
so-called "spikes" which have ligthcurves similar to the short GRBs, but
are, most probably, associated to cosmic rays hitting the ACS detector \citep{rau05}. Association of the spikes with the cosmic ray events was
questioned by \cite{acsnicolas}, who argued that the cosmic ray interactions in the SPI-ACS
could not manifest themselves as events with high count
statistics. Instead, \cite{acsnicolas} have advanced a hypothesis of possible existence of real very short GRBs missed by the other instruments. Later,
 \cite{Minaev10} stacked the profiles of a sample of unidentified short bursts in the SPI-ACS and found a weak but recognizable afterglow of the duration of more
 than a second. They interpreted this observation as an evidence in favor of the
hypothesis of the association of the (fraction of) spikes with the real short GRBs.

Our study of the origin of the short spikes reported below, agrees with the
principal conclusion of \cite{rau05} that the spikes are events induced by
high-energy cosmic rays. At the same time, we show that the instrument effects
of the ACS, the decay of cosmic ray induced radioactivity in BGO crystals,
produces the "afterglow" of the spikes discussed by \citet{Minaev10}.  The
detailed knowledge of properties of the spikes allows us to fully characterize
them and to separate them from the real GRBs in our analysis.

\section{ The INTEGRAL/SPI-ACS }\label{theacs}

The spectrometer SPI \citep{thespi} onboard INTEGRAL \cite{theintegral} is
surrounded by an active ACS \citep{spigrb}, consisting of $91$ (only $89$
currently functional) BGO (Bismuth Germanate, $Bi_4Ge_3O_{12}$) scintillator
crystals. The SPI-ACS also serves as a large effective area (up to $\sim
1$~m$^2$) \gr\ detector with a quasi-omnidirectional field of view
\citep{acs}. The design of the ACS is such that it has (almost) no sensitivity
to the direction of \gr s.

The exact magnitude of the SPI-ACS effective area and its dependency on the
direction and the energy is not well-known. It can be investigated through
detailed simulations of the photon propagation in the detector, as it was done,
for example by \citet{theaxpm}. However, this requires a mass model of the entire
INTEGRAL satellite. An alternative method consists in making use of the events
detected simultaneously by SPI-ACS and other detectors. This approach was
exploited by \citet{acscalibr}.

The ACS data are event rates integrated over all the scintillator crystals with a
time resolution of $50$~ms.  The typical number of counts per $50$~ms time
ranges from about $3000$ to $6000$ (or even more during high Solar activity).

The unique property of the SPI-ACS data is that, contrary to other existing GRB
detectors, the readout does not rely on any trigger, so that a complete history
of the detector count rate over the 10~yr period is recorded. This opens the
possibility of an off-line search of GRBs in the compete data sample.

The main challenge for the detection of GRBs in the countrate data of SPI-ACS is
in the separation of real GRBs from other sources of variability of the signal, like
the high-energy cosmic ray hits or charged particle events from the Solar
flares. In the imaging detectors of GRBs, such separation could be done based on
the directionality of \gr s contributing to the event. In the SPI-ACS separation
of GRBs has to be done based on different selection criteria.

The burst like events are detected on top of a steady background. The nearly
constant background evolves only by a factor of two over the ten years of
operation.  The background rate increases during the periods of passage of
INTEGRAL sattellite through the Earth radiation belt and during the SPI
"annealing" phases. These periods are excluded from the analysis reported below.

The SPI-ACS data are searched for bursts in real time as part of the telescope
operations, using a dedicated INTEGRAL Burst Alert System (IBAS)
\citep{theibas}.  The SPI-ACS triggers are used in the Inter-Planetary Network,
the IPN \citep{acsipn} to derive the burst positions by the triangulation if
other satellites of the IPN also detect the burst.

Comparison of SPI-ACS to other similar scintillator type GRB detectors is given
in Table \ref{tab:other}. From this table one could see that the only GRB
detector with the effective area comparable to that of the SPI-ACS was BATSE on
board of the CGRO mission.

\begin{table*}
\begin{tabular}{l|lllll}
  & Suzaku/HXD-II WAM & CGRO/BATSE LAD & Beppo-SAX/PDS GRBM & INTEGRAL/SPI-ACS & Fermi GBM \\
  \hline
  Crystal & BGO & NaI(T1)& CsI(Na) & BGO & BGO \\
  \hline
  Energy range (keV) & 50-5000 & 20-2000 & 40-700 & $>75$ & 150-30000 \\
  \hline
  Effective area (cm2) & 800@100~keV  & 2000@100 keV & 700@200 keV  & 3000@100keV & 120@200 keV  \\
  &400@1 MeV&150@1~MeV &100@1 MeV&$\lesssim$ 8000@1MeV&120@1MeV\\
  \hline
  Time resolution	& 31.25 ms & 2 ms & 7.8 ms & 50 ms & $5\ \mu$s\\
  \hline
\end{tabular}
\caption{Comparison of SPI-ACS to other GRB detectors.}
\label{tab:other}
\end{table*}

\section{Detection of GRBs }

We use the SPI-ACS data of $1030$ revolutions of the INTEGRAL satellites, from
$20$ to $1100$ ((3230 days from 2002-12-12 to 2011-10-16)). Some
revolutions are excluded entirely because of the high solar activity or SPI
annealing. The remaining total exposure time is $2412$ days.

\subsection{ Detection algorithm}\label{sec:detection}

To identify burst-like events on top to the average background of SPI-ACS, we
build the running mean of the ACS count rate on $15$ different time scales, from
$100$~ms to $10^4$~s. At each time scale $\delta t_s$, the estimate of
background count rate is obtained from the current value of the count rate on
larger time scales, $\delta t_b>\delta t_s$. The burst-like events in the
lightcurve with the time binning $\delta t_s$ are identified as excesses in the
count rate over the background count rate estimated in this way.  Significance
of detection of the burst detection depends, in general on the time scale
$\delta t_b$ used for the background rate estimate. We consider a burst-like
event appearing on a time scale $\delta t_s<20$~s as significantly detected if
there is at least one scale $\delta t_b$ for which the significance of the
signal detection on top of the estimated background exceeds a pre-defined
threshold.  The pair $\delta t_s, \delta t_b$ for which the significance of
detection of the burst is the highest defines the "best time scales" and this
highest detection significance is considered as the burst detection
significance.

This burst detection method includes an optimization for the signal/background time
scales combination. This introduces a trial factor, which has to be taken into
account in the significance estimate. Taking this into account, we set the
detection significance threshold at a relatively high value, $6\sigma$ which
corresponds to $\approx 5\sigma$ after the account of the trial factor.

The brightest bursts affect the running-mean estimates of the background even on
the largest time scales. To correct for this effect, we perform the detection
procedure twice, excluding the bursts detected at the first stage from the
background estimation at the second stage.

The background count rate is additionally affected by Solar activity. We evaluate the Solar activity related  variability of the background via monitoring of the excess variance of the background count rate on the "best" background time scale $\delta t_b$ for every burst. The significance of the burst detection is corrected taking into account possible excess variance of the background.

\subsection{ Duration measurements }\label{burstanalysis}

The most common and straightforward definition of the burst duration is the
length of the time interval in which the burst signal is detected as an excess
over the background. With this definition, the measurement of the burst duration is
obtained via a measurement of the significance of the burst signal in each
bin of the lightcurve. The duration of the burst depends on the detection
significance threshold adopted in the analysis. This means that the burst
duration defined in this way is not necessarily the characteristics of the burst
itself, but depends on the sensitivity of the instrument. Another drawback of
this definition of the burst duration is that it could not take into account
correlations between adjacent time bins. This leads to systematic
underestimation of the duration: the burst could be not detectable in the narrow
time bins, but still be significantly detectable on several time bins scale.

An alternative technique used to estimate burst durations is Bayesian Block
analysis \citep{bblocks}. It consists in approximating the burst lightcurve with
a sequence of intervals with constant flux. At first one interval is assumed. In
each step of the iterative procedure one boundary is introduced. The iterations
stop, when no more divisions are supported by the likelihood ratio test. This
method has the advantage of not relying on an a-priori known background. This is
particularly useful when the background is rapidly changing like in BAT and
GBM. One of the disadvantages of the method is that it adopts a model of the
signal in each block (constant) which is not physically motivated. This could
lead to an underestimation of the duration in the case of long and weak bursts.

The method in which the burst duration is calculated based on significance of
detection in fixed time bins could be improved to take into account weaker
signals in several adjacent time bins. For this we run the analysis on several
different pre-defined time scales. The set of the time scales is the same as
used in the detection, see Section \ref{sec:detection}. We calculate the error
on the burst durations from the comparison of the durations obtained assuming
different detection thresholds, ($4$, $3$ and $2$ sigma). In general, durations
calculated using different binning of the lightcurve are different. We compare
the durations calculated for different binnings and choose the time scale which
provides the minimal uncertainty to measure the start and stop of the burst.

Similarly to the bayesian blocks methodology, the methods adopting binning of
the lightcurves implicitly assume constant rate within the bin. Physically
motivated models of the bursts profiles can be more successful in interpreting
the properties of the burst prompt emission light curve. For example, profiles
of the burst may be fitted with a collection of pulses
\citep{norris96}. However, more complex models the compromise generality of the
analysis.

To verify the precision of the measurement of the burst durations, we perform
an analysis based on the fixed time bins and on the Bayesian blocks and compare the
results. We find that the burst durations found using the two methods are
consistent within the uncertainties.

The duration methods discussed above are based on statistical testing of the
signal over the background. Since different instruments have different
sensitivity to the signal and different background levels, the durations
measured in this way depend on the instrument performance. A definition of the
burst duration which is less dependent on the instrument sensitivity is the
so-called $T_{90}$-duration, which is equal to the time during which $90\%$ of
the burst flux arrives.  To measure $T_{90}$ we identified the total flux within
the detected duration, and adjust the start and the stop until $5\%$ of the flux
is removed from the beginning and from the end of the burst.

\subsection{Types of burst-like events in SPI-ACS}\label{burstanalysis}

The total number of burst-like events found in our analysis is $82231$.  Most of these events are a special type of events called "short spikes" \citep{rau05}. At long time scales, variations of the SPI-ACS count rate are mostly due to the changes of Solar activity. Below we explain the algorithm used to separate the GRBs from the spikes at short time scales and from the Solar flares at long time scales.   

\subsubsection{ Solar activity }\label{solar}

The Solar activity in the SPI-ACS introduces variability of various magnitudes,
ranging from the minor flickering and slow evolution to giant flares increasing
the SPI-ACS background by factor of few for days. As mentioned above, the variability of the background due to the solar activity poses a problem for the correct calculation of significance of the detection of GRBs. Our procedure for the calculation of significance of the burst takes into account the excess variance of the background during the periods of Solar activity.

Lightcurves of some solar flares could sometimes appear similar to those of GRBs.  It is not possible to distinguish such flares
from GRBs based on the SPI-ACS data alone. To distinguish the Solar flares from
the GRBs we use the data of monitoring of the Solar activity by the solar
weather prediciton center (SWPC) \footnote{http://www.swpc.noaa.gov/} to exclude time
intervals which might be affected by the Solar activity.

\subsection{ Short spikes }\label{shortspike}

The majority of burst-like events in SPI-ACS are ``short spikes'', with typical durations of about the about
$50-100$~ms, i.e. one or two bins of the lightcurve. The rate of these events
is very high - about $30$ spikes/day \citep{rau05,nicolasspikes2}. Spikes are usually detected only in one $50$~ms bin
and very infrequently in two, which suggests that the real duration is much
smaller than the bin size. 

The nature of these events was controversial.  The analysis by \citet{rau05}
suggests that the spikes are produced following hits of the ACS by
high-energy cosmic rays. The observed flux of high-energy cosmic rays is
\citep{crfluxpamela}
\begin{equation}
  F_{CR}\simeq 10^2 \left[\frac{E_{CR}}{10\mbox{ TeV}}\right]^{-1.7}\mbox{m}^{-2}\mbox{sr}^{-1} \mbox{d}^{-1}
\end{equation}
Taking into account the effective area of ACS, the observed rate of spikes
suggests that the events are produced by cosmic rays with energies above
$\sim 30$~TeV.

Alternatively, the spikes (or a fraction of them) could be real GRBs which
escape detection by other instruments. This could arise because of large
effective area of ACS at the energies above $\sim 1$~MeV. If the spikes are real
GRBs, they should have very hard intrinsic spectrum, to avoid detection by other
instruments.  The hard and very short bursts would be a phenomenon of particular
interest. A distinct population of bursts with similar properties might have
been found in the Swift/BAT, BATSE and Konus Wind burst samples
\citep{cline99,vsb}.

We dedicate the next Section to study of the Short Spike properties.

\section{ The nature of the short spikes }\label{spikenature}

The properties of the short spikes reveal a remarkable stability over the whole
mission span (10 years of data). While the background is slowly changing with
the solar cycle, the sensitivity-corrected spike rate remains constant. This
observation makes a Solar origin of spikes unlikely, unless there is a
high-energy solar phenomenum not modulated with solar cycles. If cosmic rays are
resposible for the spikes, however, they should not be affected by the solar
activity and thus have energy higher than $\gtrsim 10$~GeV, where the Solar
modulation effect vanishes \citep{crmodulorig,crmodulbess}

To characterize the time profile of the  spikes we stack the lightcurves of all the
very short bursts detected by SPI-ACS (shorted then $0.5$~s) shifted in time to match
the bin with the highest countrate.

\begin{figure}
  \includegraphics[width=\columnwidth,angle=0]{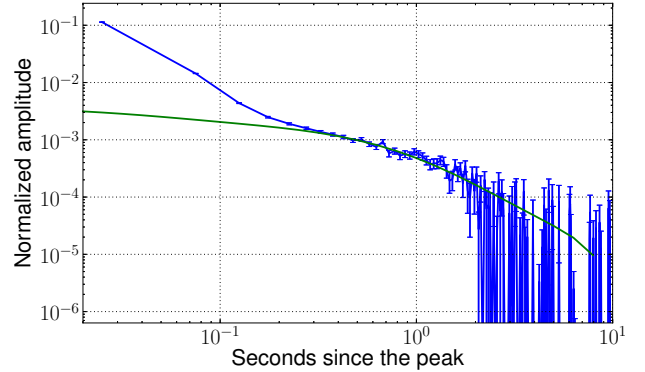}
  \caption{The stacked profile of the spikes (blue data points).  The extended tail is fitted with the decay of residual cosmic ray induced radioactivity of the BGO found from FLUKA simulations (green curve).  \label{fig:afterglow}}
\end{figure}

The stacking analysis reveals a clearly recognizable weak ($0.1~\%$ of the total
number of counts or just a few counts in a single spike on average) afterglow
lasting for almost $3$ seconds (Figure \ref{fig:afterglow}). From
this figure one
could see that the "prompt" phase of the spike is characterized
by an exponential rise / exponential decay profile. The rise and decay time are
constrained to be shorter than $\sim 25$~ms, half of the time bin of the ACS
lightcurve. The overall event number and count statistics in the stacked sample
is largely dominated by spikes, so that the presence of confirmed short GRBs or
SGR flares in the sample does not affect the average spike profile found from
the stacking analysis. To verify that the long tail of the stacked profile is
not due to the presence of a sub-population of rare real GRBs in the stacked
sample, we studied distribution of countrates in individual events in the time bins of the tail. This distribution follows a Gaussian with the center displaced from zero. If the tail of the profile would be due to the presence of a small
number of real GRBs, the distribution would be instead a Gaussian centered at
zero, with a small excess at positive count rates. We have also verified that
the profiles of individual brightest spikes follow closely the averaged stacked
profile, with a clearly visible tail.

\citet{rau05} suggested that the spikes are produced following energy deposition
in the ACS by a high-energy cosmic ray.  The cosmic ray initiates a shower of
secondary particles through pair production and bremmstralung and then
propagating further to the $Ge$ detectors of SPI and saturating them. One of the arguments presented by \citet{rau05} in favor of interpretating of the spikes as events produced by the high-energy cosmic rays is the correlation of the spikes with events of saturation of the SPI detector.  \citet{rau05} noticed that
saturation of at least one $Ge$ detectors of the SPI often (but not always)
accompanies a short spike and that saturation of more than one SPI detector is
always accompanied by a spike-like increase in SPI-ACS count-rate.

Our analysis of the correlation between the spikes and the saturation of SPI
detectors does not fully confirm the observation of \citet{rau05}. Although we
find a significant correlation between the spikes and SPI saturation moments
(1600 coincidence events with expected 40), we find that saturations of more
than one SPI detector are not always accompanied by spikes.

In particular, the saturation of all the SPI detectors is always accompanied by a
gap in the SPI-ACS data as well, so that no spikes could be recorded. Although
there is a significant correlation between spikes and "cosmic-ray track-like"
saturation events in SPI (more than $3$ detectors in a row are saturated), some
$60~\%$ of the track-like events are not accompanied by spikes. The correlation
is still weaker in the saturation events which do not have track-like
appearance. Only 5\% of those events correlate with the spikes.

Problems of this hypothesis were highlighted by \citet{nicolasspikes2}.
\citet{acsnicolas} have noticed that all the particles from a cosmic ray induced
shower in ACS would be counted as just one particle, because of the finite time
resolution of the ACS counting electronics ($0.6\ \mu$s). Indeed, the shower
particles travel with a speed close to the speed of light. The transit time of
the ACS is much shorter than the time resolution of the detector. This means
that the cosmic ray induced shower could not produce a spike in the ACS count
rate. The shower particles also could not be responsible for the
second-long afterglow shown in Fig. \ref{fig:afterglow}, as highlighted by \citet{Minaev10}.

A potential way out of this problem is that the cosmic ray induced
shower in ACS is able to excite long-lived phosphorescence of the
detector. In this case the phosphorescence signal persists on longer
time scale and could be time-resolved by the ACS front-end electronics. This effect
is actually observed in another detector on board of INTEGRAL, PICsIT, in which
strong ($\sim2500$~counts~s$^{-1}$) and short ($\lesssim 170$~ms) spikes are
associated with cosmic ray induced showers \citep{segreto03}. Time-tagging of
photons and energy resolution of PICSIT allow identification of the spikes as
the phosphorescence of the detectors following the cosmic ray hits.

SPI-ACS uses undoped BGO scintillators, featuring very rapid deexcitation - with
a characteristic times of about $60-300$~ns  followed by weak,
$0.005\%$, residual phosphorescence on the time scale of $3$~ms. If the energy
deposit by a cosmic ray is large enough, even weak residual phosphorescence
signal could be above the trigger threshold of the ACS front-end electronics. In
this case, the detector will continue to count events on the time scale up to
the $\sim$~ms following the CR hit. This would explain the appearance of a spike
in the ACS count rate as suggested by \citet{rau05}.

However, prolonged phosphorescence of BGO crystal could not be responsible for
the spike afterglows shown in Fig. \ref{fig:afterglow}, because persistence of the phosphorescence signal on the $\sim$second time scale would produce a flat constantly high countrate $R\sim 50\mbox{ ms}/0.6\ \mu\mbox{s} \simeq 80000$~counts/bin on this time scale, rather than "spike-with-a-tail" lightcurve. Instead, the tail could be attributed to another feature of the BGO crystal, the induced radioactivity. Activation of the BGO crystals with the cosmic ray protons is known to contribute to the measured background \citep{bgoact1997ESA}.  We have investigated the radioactivity induced in the BGO crystals by high-energy cosmic ray passage using FLUKA Monte-Carlo code \citep{fluka1,fluka2}. These simulations reveal that the main radioactive decay channel is the decay of ${}^{8}$Li with the half-life time $838$~ms. There is
also a number of other isotopes with comparable decay times although lower abundances. A fit of the observed time profiles of the extended tails of the spikes with the time profiles of decay of the cosmic ray induced radioactivity
found from FLUKA simulations is shown in Fig.  \ref{fig:afterglow}. One could see that the time profile of the tail is well explained by the exponential decay of the induced radioactivity in the BGO crystal.

To summarize, the observed properties of the spikes are consistent with the
assumption that they are produced by high-energy cosmic rays passing through
the ACS and inducing phosphorescence and radioactivity in the BGO crystals.

Knowledge of the physical origin of the spikes enables us to efficiently reject
them in the analysis of the GRB sample. All the spikes are expected to have
identical time profiles, with the overall normalization determined by the energy
of the primary cosmic ray. Rejecting all the events which have the time profiles
consistent with that of the template spike profile shown in Fig. \ref{fig:afterglow}, will remove most of the spikes, while
suppressing only a small fraction of the real GRBs which occasionally have time
profiles similar to the spikes (a pronounced peak followed by a tail with fixed
fractional amplitude and decay time). Taking this into account, we compute the
likelihood for each event to be a spike rather than real GRB. We calibrate this
likelihood with simulations and choose the threshold to reject the spikes in the
real data.

\section{Properties of GRBs detected by SPI-ACS}\label{sec:grbs}

\subsection{GRB durations}

\begin{figure}
\includegraphics[width=\columnwidth,angle=0]{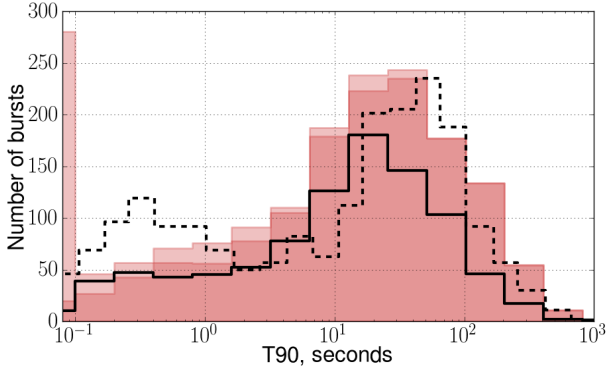}
\caption{Distributions of the $T_{90}$ durations. Dark and light shadings of the
  filled region correspond to 5 and 3 sigma confidence level for the event not being a spike. Solid black curve is for the confirmed bursts. The distribution of durations of GRBs from BATSE 4B GRB sample, scaled to match the normalization of the ACS sample, is shown by the black dashed line.\label{fig:durations}}
\end{figure}

\begin{figure}
\includegraphics[width=\columnwidth,angle=0]{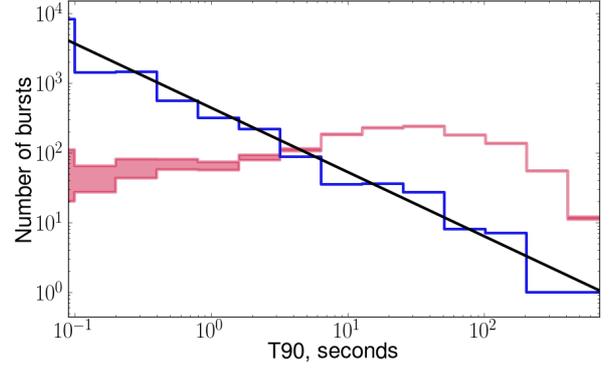}
\caption{Distribution of GRB durations (shaded region) compared to the
  distribution of durations of spikes (blue
  histogram).\label{fig:durations_log}}
\end{figure}

Rejection of spikes and Solar flare events from the overall sample of burst-like
events in SPI-ACS leaves a sample of 1416 GRB lightcurves
available for the timing analysis. Figure \ref{fig:durations} shows the
distribution of durations $T_{90}$ for these GRBs. This distribution is compared
to the distribution of durations of the spikes shown by the blue histogram in Fig. \ref{fig:durations_log}.

One could see that the spikes dominate the total number of events with durations
up to several seconds. However, the characteristic time profile of the spikes
allows their efficient rejection even on the shortest time scales. In
particular, a significant number of GRBs with durations down to $\sim
100-150$~ms is detected in the SPI-ACS data. Such bursts are identified as
events for which the lightcurve is not consistent with the spike time profile. Two shaded histograms in Fig. \ref{fig:durations} show two different confidence levels for an events not to be spikes.  One could see that reducing the acceptance threshold leads to a sharp increase of the number of GRB-like events
on the shortest time scale where difference between the spikes and real GRBs is
more difficult to characterize. Taking this into account, we restrict our attention to GRBs with durations longer than 100~ms in the following.

Distribution of durations of the spikes follows a powerlaw on large time
scales. This is explained by the fact that the span of the time interval during
which the cosmic ray induced radioactivity produces a detectable signal in the
BGO crystal scales proportionally to the overall number of the radioactive atoms
which is in turn proportional to the energy of the primary cosmic ray. Taking
into account that the spectrum of cosmic rays follows a powerlaw, one could find
that the distribution of durations of the spikes should also follow a powerlaw.
Amplitude of statistical fluctuations of the spike duration distribution
determines an upper bound on the possible number of real GRBs in each duration
time bin. This upper limit is shown in Figure \ref{fig:durations_log} as an upper
boundary of the pink-shaded region.  Lower boundary of the pink shaded region is
the distribution of the GRB-like events calculated with tight rejection
threshold for the spikes, shown in Figure \ref{fig:durations}.

Real GRBs are expected to produce signals not only in ACS, but also
simultaneously in other GRB detectors. To verify the quality of our procedure of
selection of GRB events from the set of burst-like events in SPI-ACS, we have
searched for the events occurring in coincidence with triggers of other GRB
detectors.  Table \ref{table:assoc} summarizes the result of our search. The
number of events found in coincidence with triggers of other detectors is
comparable to the number of events selected based on the SPI-ACS data alone.
Slightly larger number of events found using SPI-ACS alone is readily explained
by the large effetive area of ACS. In our search of events occurring in
coincidence with triggers of other detectors, we have adjusted the search time
window (within which the events are declared to be coincident) in such a way
that the we expect $1$ random coincidence.

\begin{table*}
\begin{tabular}{ l | *{7}{p{2.cm}}}
& Konus Wind & Suzaku WAM & Fermi GBM & RHESSI & Swift/BAT & IPN & Fermi LAT \\
\hline
GRB & 461        & 452        & 292       & 145    & 127       & 21  & 16 \\
\end{tabular}
\caption{SPI-ACS triggers detected simultaneously by other means}
\label{table:assoc}
\end{table*}

The black solid histogram in Figure \ref{fig:durations} shows the distribution of
durations of events occurring in coincidence with the triggers from other
detectors. One could see that this distribution is consistent with that found
based on the analysis of the SPI-ACS data alone, although with somewhat lower
statistics.

Although the presence of the short spikes reduces the sensitivity of the SPI-ACS to the shortest ($< 100$~ms) bursts, the number of short bursts with longer durations (100~ms - 2s) is rather well constrained. The fact that distribution of durations of the independently detected bursts is so similar to the distribution of the confirmed burst durations (Figure \ref{fig:durations}), shows that our spike filtering technique efficiently rejects most of the spikes and retains most of the bursts with durations longer than two ACS time bins.

\subsection{GRB peak fluxes}

\begin{figure}
\includegraphics[width=\columnwidth,angle=0]{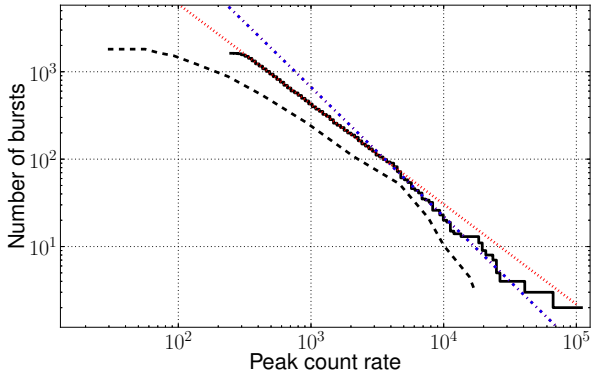}
\caption{Peak count rate distribution of the GRB candidate events. Powerlaw fits above and below the
  break at $4000$~counts/s are shown with dotted lines. Distribution of the GRB peak fluxes as
  measured by BATSE (rescaled) is shown by the black dashed line.}
\label{fig:peakcnt_distr}
\end{figure}

To verify the efficiency of rejection of spikes based on the template profile,
we study the distribution of peak countrates in the remaining GRB sample.  The
peak countrate measured by SPI-ACS provides an estimate of the peak flux of the
GRB.  SPI-ACS detector could not measure the peak flux of each detected GRB,
because its effective area depends on the (in general, unknown) arrival
direction of GRB photons. However, GRBs arrive isotropically, so that the
average peak flux distribution could be measured for all the GRB population,
using a known angle-averaged effecitve area of ACS. This distirbutiton is shown
in Figure \ref{fig:peakcnt_distr}. One could see that the distribution has the
form of a broken powerlaw with slopes $-1.46\pm0.04$ at high countrates and
$-1.12\pm0.01$ at the countrates below 4000. This count rate corresponds to a
GRB peak flux $\sim 10$~ph/cm$^2$s in the energy band above 80~keV or,
equivalently, to an energy flux $5\times 10^{-6}$~erg/cm$^2$~s. For comparison,
the distribution of the GRB peak fluxes as measured by BATSE
\citep{batsegrbintensity} is shown at the same figure. It is scaled for the
convenience of the presentation. Notably, both the break and the slopes in the
BATSE peak flux distribution are close to the one we observe.

\subsection{Characteristic time scale of long GRBs}\label{burstanalysis}

We have investigated the variability patterns of the GRB lightcurves, using
conventional time series analysis tools, such as power spectral density (PSD)
and structure funciton analysis. These tools are well suited for the analysis of
long GRBs, in which the dynamic range of time-scales / frequencies for each GRB
event is large enough.

We use the first order structure function to quantify the burst temporal
properties. The structure function analysis is analogous to the autocorrelation
function (ACF) analysis, successfully used by a number of authors in
investigations of the GRB variability \citep{grbacfbimodal,acsshots}. We prefer the structure
function analysis to the PSD analysis which is better suited for the study of
variability of persistent, rather than transient sources, because of the nature
of the Fourier transform.

The structure function analysis could be used to identify a characteristic
variability time scale, which appears as a localized "deep" feature superimposed
on the powerlaw type behavior. We use this property of the structure function to
investigate the possibile existence of a characteristic variability time
scale of GRBs. The existence of such a timescale in the $T_{var}\simeq 1$~s range
was claimed by \cite{grbpsd} based on the PSD analysis of a set of BATSE GRBs.

\begin{figure}
\includegraphics[width=\columnwidth,angle=0]{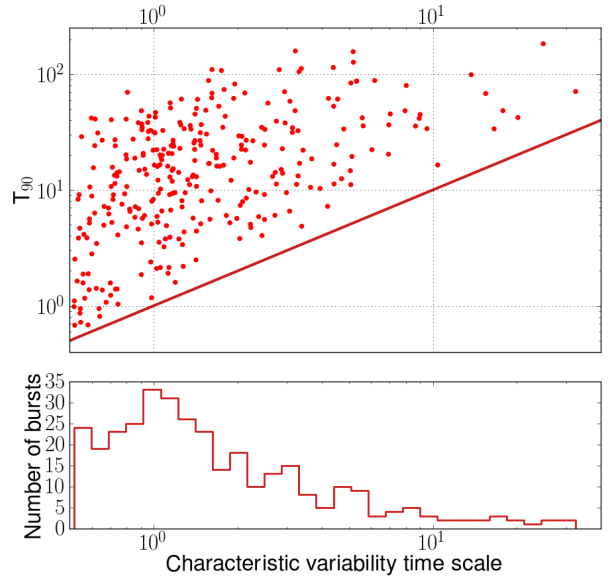}
\caption{One the top panel: the first characteristic time scale vs
    duration ($T_{90}$) for long GRBs\label{fig:chartime_dur}. On the bottom
    panel: the distribution of the first characteristic time scales.}
\end{figure}

Our structurre function analysis confirms the result of \cite{grbpsd}.  Figure
\ref{fig:chartime_dur} shows the characteristic time scales of long GRBs found
form the structure function analysis and plotted as a function of the GRB
duration. One could see that large fraction of GRBs has the characteristic time
scale $T_{var}=1\pm1$~s.

\section{Discussion}

Our analysis of the sample of GRBs detected by SPI-ACS shows that the
distribution of GRB durations shown in Figure \ref{fig:durations} has two
components. Similarly to BATSE duration distribution, it has broad bump
corresponding to the long GRB population. The typical duration of long GRBs in
SPI-ACS ($\simeq 20$~s) is somewhat shorter than in BATSE ($\simeq 50$~s).  This
is in agreement with the previous studies of the SPI-ACS bursts performed by
\citet{rau05}. The difference in the typical duration could be attributed to the
difference in the energy ranges in which BATSE and SPI-ACS are sensitive. The
dependence of the duration distribution on the energy range was
previously observed in Swift/BAT, BATSE, Beppo-SAX and HETE data
\citep{fenimore95,bat2ndcat}. Longer bursts are known to have softer
spectra. Besides, the spectra of individual GRBs have hard-to-soft evolution so
that the burst duration grows with decreasing photon energy.

Contrary to BATSE burst duration distribution, the distribution of durations of
SPI-ACS bursts, shown in Figures \ref{fig:durations} and
\ref{fig:durations_log}, does not reveal short bursts concentrated around a
characteristic time scale. Instead, the burst duration distribution follows a
powerlaw at short time scales $T_{90}\lesssim 3$~s. The powerlaw extends to the
shortest time scale accessible to the ACS ($\simeq 100$~ms) and, most probably,
to the longer time scales, beyond $\sim 3$~s.

This might be either due to the instrumental effects or might reflect the GRB
physics. An instrumental effect could arise e.g. because of the difference of
sensitivity between BATSE and SPI/ACS. If short GRBs typically have extended
tails, which were not detected by BATSE, but are visible to SPI/ACS, part of the
bursts which might have been responsible to the short burst duration
distribution bump at 0.3~s, might appear in SPI/ACS as longer bursts in the
$\sim 3$~s duration range. Otherwise, the difference between the short and the
long bursts may not be fully characterized by the criterion $T_{90}\lessgtr
2$~s, as suggested by BATSE distribution. Separation in the hardness and
spectral lags have to be used to establish the separation between the two main
varieties of the GRBs. As discussed by \citet{Zhang11_grbstatus}, it may be
natural to adopt a different classification scheme: to Type I ("compact star
GRBs") and Type II ("massive star GRBs") bursts.

Although our analysis does not reveal a separate population of very short bursts
with the durations $T_{90}<100$~ms, discussed by \citet{vsb}, we could not
completely rule out the existence of such sub-population. Such very short bursts
could be easily confused with the cosmic ray induced spikes in ACS.  To
distinguish real bursts from the cosmic-ray induced spikes a more advanced
analysis based on the simultanenous detections by the other intruments,
e.g. Suzaku/WAM, has to be done.

The sample of the SPI-ACS bursts, presented in this work, is, of course,
  subject to observational biases related to the detection efficiency of our
  method. It was shown that in the case of BATSE specific detection efficiency
  issues have significant effect on the burst duration distribution, increasing
  the fraction of short bursts to about 40\% \citep{lee95}. Unfortunately, low
  number of measurable burst characteristics accessible in SPI-ACS (limited to
  timing properties of lightcurves in a single energy band) make similar
  analysis impossible in the case of SPI-ACS. A detailed study of the SPI-ACS
  GRBs detected simultaneously with other satellites may help in recovering the
  intrinsic distributions of the GRB properties. Alternatively, selection biases
  might be assessed via a consideration of different trigger criteria. It has
  been shown that a sample of the short bursts may be better probed with the
  time-to-spill flux-based rather the traditional fluence-based trigger
  \citep{norris84}.

The second result of our work is the identification of a characteristic time
scale of variability of long GRBs, see Figure \ref{fig:chartime_dur}. This time
scale can be roughly attributed to the typical separation between the
  individual pulses or activity periods of GRBs. It is interesting to note that
the characteristic time scale of long GRBs coinsides with the time scale at
which a break in the distribution of GRB durations is observed (see Figure
\ref{fig:durations_log}). This might indicate that many long GRBs are sequences
of $\sim 1$~s pulses, with the overall burst duration determined by the overall
number of pulses. The "pileup" in the distribution of GRB durations, at the time
scale $\sim 20$~s shows that long GRBs typically consist of $\sim 10-20$ pulses.

The majority of GRB with measured redshifts are situated at $z\sim 1-2$
\citep{swiftgrbcompletez}. This implies that the characteristic time scale in
the GRB rest frame is of the order of $0.5$~s.

Our measurement of the characteristic pulse scale is robust but
  simplistic. Extensive studies of the individual pulse properties were
  performed by a number of groups \cite{norris96,hakkila09,hakkila11}. In their
  pioneering work \citet{norris96}, matched the observed lightcurves by a
  combination of model pulses. Notably, their result - the characteristic time
  scale of the order of a second, is close to the one we obtained with
  model-independed structure function analysis.

Results of \citep{norris96} and further studies
  \citep[e.g.][]{hakkila08,ryde02} also indicated that the pulse properties
  (including characteristic timescales) are correlated. Since we use a wide range
  of the peak fluxes, it may should have an effect on our measurement of the
  characteristic $1$~second scale. However, this time scale is readily
  detectable in the SPI-ACS data even after smoothing introduced by ignoring the
  correlations. The robustnes of the detection supports the idea that it should
  be associated with a real physical characteristic of the bursts.

The distinct characteristic time scale of the order of a second might be
associated with the variability of the GRB central engine. In this case the
remarkable persistency of the this time scale seems to be naturally attributed
to the similarity of the progenitors \citep{grbpsd}. Alternatively, it can be
related to the size of the emission region.  Further investigation of the nature
of the characteristic time scale, exploiting the rich timing data of the
SPI-ACS, is required to clarify its origin.

\section{Acknoledgement}

We would like to thank Nicolas Produit for the useful discussions of the subject
and for the comments on the manuscript.

\bibliographystyle{aa}

 \label{lastpage}

\end{document}